\documentclass[prb,aps]{revtex4}
\usepackage{epsfig,epsf}
\usepackage{epic,eepic}
\usepackage{color,pstcol}
\usepackage{graphicx}
\begin{document}
\title{ Nonlocal transport through multiterminal diffusive superconducting nanostructures}
\author{F. S. Bergeret}
\affiliation{Departamento de F\'{\i}sica Te\'orica de la Materia Condensada C-V, Facultad de Ciencias, Universidad Aut\'onoma de Madrid, E-28049 Madrid, Spain.\\ 
Centro de F\'{i}sica de Materiales (CFM), Centro Mixto CSIC-UPV/EHU, Edificio Korta, Avenida de Tolosa 72, 20018 San Sebasti\'{a}n, Spain.\\
Donostia International Physics Center (DIPC), Manuel de Lardizábal 4, 20018 San Sebasti\'{a}n, Spain.}
\author{A. Levy Yeyati}
\affiliation{Departamento de F\'{\i}sica Te\'orica de la Materia
Condensada C-V, Facultad de Ciencias, Universidad Aut\'onoma de Madrid,
E-28049 Madrid, Spain}
\begin{abstract}
Motivated by recent experiments on nonlocal transport through  multiterminal 
superconducting hybrid structures, we present self-consistent calculations 
based on quasiclassical Green's functions for the order parameter, currents 
and voltages in a system consisting of a  diffusive superconductor connected to two normal and one superconducting  electrodes. We investigate non-equilibrium 
effects for different biasing conditions  corresponding to measurements of 
the nonlocal conductance and of the nonlocal resistance.
It is shown that while  the nonlocal conductance does not  
change its sign, this change might be observed in a nonlocal resistance 
measurement for certain parameter range. The change of sign of the 
nonlocal signal takes places at a voltage of the order of the self-consistent 
gap of the superconducting region. We show that this 
is not related to the nonlocal Andreev processes but rather to
non-equilibrium effects. We finally discuss the case of four terminal
measurements and demonstrate that a change of sign in the nonlocal
resistance appears when the current injected into the superconductor
exceeds a critical value. The connection to the existing experiments is discussed.
\end{abstract}
\maketitle

\section{Introduction}
The possibility to create and control entangled electron pairs from a superconductor has renewed the interest in 
transport through superconductor-normal metal hybrid structures \cite{beckmann,delft,chan,beckmann2,basel}. The basic idea 
is to exploit the long-range coherence of the Andreev reflection\cite{andreev}, in order to couple two spatially separated 
normal electrodes connected to a superconducting region within a distance of the order of the superconducting coherence 
length $\xi_S$. An Andreev processes that takes place at two different interfaces is called a "crossed"' Andreev 
reflection (CAR). As suggested in Ref. \cite{byers} spatial correlations can be probed by nonlocal transport experiments.  
A typical  setup for detection of CAR processes  consists of  a grounded superconducting region (S)   connected to  normal 
electrodes (N). The information on such processes would be encoded, for instance, in the voltage which is measured in one of the S/N interfaces when a current is injected through the other one. Besides CAR processes, individual electrons can also tunnel across the superconductor. This normal tunneling has been called ``elastic cotunneling'' (EC). 
CAR and EC contributions to the non-local conductance have opposite signs, and 
in the lowest order of tunneling  cancel each other\cite{falci}.  For higher orders in the tunneling the EC  dominates over CAR \cite{regis} and the nonlocal signal becomes finite. Surprisingly a change of sign in the nonlocal resistance and conductance was reported as a function of the local voltage for a NSN   layered structure\cite{delft} and a FSF multiterminal structure\cite{beckmann2} (F denotes a ferromagnetic metal).  According to Ref.\cite{delft}  by low (high) voltages EC (CAR) processes dominate the nonlocal transport. Similar  behavior was  observed in Ref.\cite{beckmann2} for samples with high S/F barrier resistance. The latter experiment also showed dominance of CAR processes at low voltages for samples with higher interface transparency.  These  experimental results have 
lead to   several theoretical works \cite{regis,kalenkov,morten,golubov,golubev,natphys,zaikin09,prb09}, which attempt to find a microscopic description for those observations.  However, up to now theories based on non-interacting models could not explain the change of sign of the non-local conductance \cite{kalenkov,morten,golubov,golubev}.  For a layered NSN structure, as the one of Ref. \cite{delft},  the change of sign  of the non-local conductance has  been  explained  by taking into account interaction of the conducting electrons with their electromagnetic environment\cite{natphys}. The later description is valid in the tunneling limit, thus the observation of negative non-local conductances in the case of good interface 
transparencies remains without microscopic explanation yet.  
 
In principle, a  description of the non-local transport in terms of CAR and EC is only valid in the tunneling limit. In this case the system is in a quasi-equilibrium state, {\it i.e.}  the current (or the corresponding bias voltage) is much smaller than its critical value and the distribution function of quasiparticles is the equilibrium one. This  assumption has been made in  most of  the theoretical works mentioned above. In particular  the superconducting gap was assumed to have the bulk value, {\it i.e.}   the superconducting order parameter was not affected  by the possible deviation of the distribution function from its equilibrium value.  In some experiments though,  this is not the case. For example in   Refs. \cite{chan,basel,beckmann} the   transparencies of the S/N interfaces are not necessarily low, and  in Ref. \cite{chan}  the current injected into the S region reached its  critical  value. Thus, for a proper description of these experiments one needs to go beyond the quasi-equilibrium approach.  A first attempt was done in Refs\cite{prb09,zaikin09}. It was shown that non-equilibrium effects may play a crucial role on the non-local transport properties. In particular, in Ref. \cite{prb09} self-consistent calculations 
based on  a two dimensional tight binding model were implemented. It was shown that far from 
the quasiequilibrium regime the nonlocal transport cannot longer be described 
in simple terms of EC and CAR processes. For some set of parameters,  a change 
of sign in the non-local resistance was obtained.  This change of sign  is not  related to the predominance of 
CAR but rather to the possibility of having a negative local conductance at the interface where the current is injected. 
In Ref.\cite{zaikin09} a non-monotonic behavior of the non-local resistance as 
a function of the temperature was obtained, which resembles the observations of Ref. \cite{chan}.  However, the non-local 
resistance as a function of the injected current or bias voltage was not investigated in that work.

In the above mentioned theoretical works, one computes the non-local conductance in a three terminal device. In other 
words, one assumes that one of the normal terminals is biased to a voltage $V_L$, while the second normal  terminal is 
grounded.  In this way one determines the current $I_R$ flowing into the latter terminal and computes the non-local 
conductance $G_{nl}=dI_R/dV_L$.  Experimentally, however, it is simpler to fix the injected current through one of the 
normal terminals and measure the voltage induced at the second normal terminal where no current is flowing. Thus, the 
measured quantity is the non-local resistance. 
Moreover, some experiments were performed in a  multi-terminal geometry \cite{chan}.   
  
In  this paper we present a complete self-consistent theory for 
the nonlocal transport through a diffusive superconducting region connected to 
several normal electrodes. We calculate  both, the non-local  conductance and 
resistance  with the help of the quasiclassical 
Green's functions (GF) approach.  In a first part we concentrate on a three terminal device, where the current is injected from a normal electrode $N_L$  into the superconducting region $S$,  maintaining the second normal electrode $N_R$ grounded. We determine the self-consistent gap,  the current flowing into $N_R$ and compute the non-local conductance $G_{nl}$.  We show that our model, as in previous works, predicts no change of sign for $G_{nl}$. In a second part we 
consider again a three terminal device, but now we assume that no current is 
flowing at the $S/N_R$ interface terminal. We then determine the self-consistent gap, the   voltage induced in $N_R$, 
and determine the non-local resistance 
$R_{nl}$.  For certain range of parameters we obtain a change of sign $R_{nl}$ 
due to the appearance of a negative local conductance at the $N_L/S$ interface. 
Finally, we considered a four terminal setup. Again we assume that the current 
through the $S/N_R$ interface is zero, but now we determine the voltage induced in the $N_R$ electrode measured with respect to the end of the superconductor 
in which no current is flowing. In this case we  obtain a change of sign as in 
the experiment of  Ref. \cite{chan}. As we show below the origin of this change of sign is not  due to a negative 
local conductance, but to the non-equilibrium distribution created in the superconductor by the injected current, which 
eventually leads to a transition into the normal state.

The rest of the paper is organized as follows: in section II we
introduce the model and basic equations used throughout the paper, section
III and IV are devoted to analyze the results for the three terminal
situation (we discuss the conductance measurement conditions in 
section III and the resistance measurement case in section IV), 
and finally in section V we analyze the four terminal case. Some 
concluding remarks are given in section VI. 
\begin{figure}[t]
\centerline{\includegraphics[scale=.35]{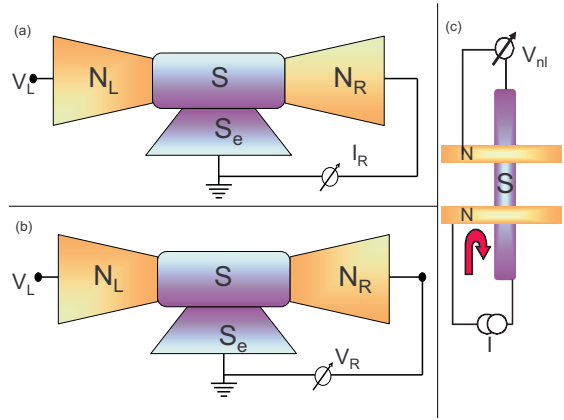}}
\caption{(Color online.) Sketch of a three-terminal device   for (a) measurement of  the non-local conductance and  (b) measurement of the  non-local resistance. (c) Typical experimental setup for measurement of the non-local resistance. 
\label{geometry}
}
\end{figure}
\section{The Model and basic equations}

A typical experimental setup for the measurement of non-local transport properties  is shown in Fig. 1c. On top of a nanoscale superconducting wire one 
places several (in our case two) normal wires.  A current is injected  from one of the normal electrodes  and  flows in the direction shown in the figure.   The non-local resistance  is then obtained by measuring the  potential difference between the other  normal electrode and the end of the superconducting wire  through which no current is flowing.  A strong enough  non-local signal is measured when the distance between the normal leads does not exceed much  the coherence length of the superconductor.

A self-consistent computation of the spatial variation of the  gap and currents in such a device is a formidable task which we will not address here. 
Nevertheless, the mean features of the system can be described by considering the geometry shown in Figs. 1a and 1b.
A superconducting region $S$ is connected to two normal electrodes N$_L$ and N$_R$ and to a superconducting electrode $S_e$, made  of the same superconducting material as $S$. The three interfaces will be described by characteristic energies  $\epsilon_L$,$\epsilon_R$ and $\epsilon_S$, defined below. 
As we are not interested in the spatial variation of the non-local  correlations, 
we simplify the problem by assuming that the central superconducting region $S$ has dimensions  smaller than the superconducting coherence length, which in the diffusive limit is given by $\xi_S=\sqrt{D/\Delta}$.  Here $\Delta$ is the superconducting gap and $D$ is the diffusion coefficient.  Thus,  we may assume that the order parameter and the non-equilibrium distribution are uniform over  $S$ \footnote{ For low temperatures and voltages, non-local correlations are governed by EC and CAR processes, {\it i.e.} their characteristic decay length is $\xi_S$, as shown in Ref.\cite{feinberg}. For higher temperatures or currents  charge imbalance effects may become important.  The length controlling the charge imbalance, though,   is larger than $\xi_S$, and therefore our assumption remains valid}. 

 In order to determine either  the non-local conductance (Fig. 1a) or   resistance (Fig. 1b) we need  to calculate the current density ${\bf j}$ and the self-consistent order parameter  $\Delta$ in the $S$ region. 
As one can  see from our results  below, we always find a stationary current state. This behavior is distinctive for junctions consisting of  a mesoscopic superconductor  in contact with a bulk one\cite{volkov}.
  Thus, ${\bf j}$ and $\Delta$  can be expressed  in terms of  the Keldysh component of the matrix GF $\check g$   
\begin{eqnarray}
\Delta&=&\frac{\lambda}{4}\int d\epsilon  \hat g^K_{12} \label{delta}\\{\bf j}&=&\frac{1}{8eG_N}\int d\epsilon{\rm Tr}\left\{\hat\tau_3\check g\nabla\check g\right\}^K\label{current}\; ,    
\end{eqnarray}
where $\lambda$ is the BCS  coupling constant which determines the critical temperature, and $R_N$ is the normal state resistance of the $S$ region.
The function $\check g$ is a 4$\times$4 matrix in the Nambu$\otimes$Keldysh space with the usual structure
\begin{equation}
\label{GF4}
\check g = \left( \begin{array}{cc}
\hat g^{R} & \hat g^K \\
0  &  \hat g^{A}
   \end{array} \right) ,
\end{equation}
while $\hat g$ are  2$\times$2 matrices in Nambu space.
In the diffusive limit these functions are the solutions  of  the Usadel equation\cite{usadel}
\begin{equation}
-D\nabla(\check g\nabla\check g)-i\epsilon\left[\hat\tau_3,\check g\right]-i\left[\check \Delta,\check g\right]=-i\left[\check\Sigma_{in},\check g\right]\; \label{usadel},
\end{equation}
supplemented by the normalization condition $\check g^2=\check{1}$. Here 
 $\check \Sigma_{in}$ is the self-energy term describing inelastic 
processes. 
In the time relaxation approach  $\check \Sigma_{in}$ is proportional to  $1/\tau_{in}$, where $\tau_{in}$ is 
the inelastic relaxation time. We will assume that $1/\tau_{in}$ is the smallest energy scale and neglect this term.
At the interfaces with the electrodes 
we    use the Kupryianov-Lukichev boundary conditions\cite{kl}
\begin{equation}
D\check g\nabla\check g|_{\bf n}=\epsilon_id\left[\check g,\check g_i\right]\; ,
\label{kl}
\end{equation}
where $\check g_i$ are the GF of the electrodes($i=L,R,S_e$), $\epsilon_i=\epsilon_{Th}/2r_B$, $\epsilon_{Th}=D/d^2$ is the Thouless energy, $r_B=R_{Bi}/R_N$,   $R_{bi}$ is the ith barrier resistance per unit area, and  ${\bf n}$ denotes a unit vector normal to the interface.
We assume that the GFs of the electrodes  remain unchanged and equal to the bulk values, i.e.   
$\hat g^{R(A)}=\pm \hat \tau_3$ in the normal leads, and    $\hat g^{R(A)}_S=g^{R(A)}_{BCS} \hat \tau_3+f^{R(A)}_{BCS}i\hat \tau_2$ in the superconductor electrode, where $g^{R(A)}_{BCS}=\epsilon/\sqrt{(\epsilon\pm i\eta)^2-\Delta_0^2}$ and $f^{R(A)}_{BCS}=\Delta_0/\sqrt{(\epsilon\pm i\eta)^2-\Delta_0^2}$. While the Keldysh components are given by 
\begin{equation}
\hat g^K_i=\hat g^R_i\hat F_i-\hat F_i\hat g^A\label{GFK}
\end{equation}
where 
\[
\hat F_i=F_{i+}\hat\tau_0+F_{i-}\hat\tau_3, 
\]
  $V_i$ is the voltage in electrode $i$ , and $F_{i\pm}=\frac{1}{2}\left[\tanh(\frac{\epsilon+eV_i}{2T})\pm\tanh(\frac{\epsilon-eV_i}{2T})\right]$ . We also assume that  $V_{S_e}=0$.
In principle the boundary conditions Eq. (\ref{kl}) are valid for low transmitting interfaces. In the present work we consider that the
interface transparencies may vary in the range $10^{-3}-10^{-1}$ for which Eq. (\ref{kl}) is sufficiently reliable.

 With the help of Eq. (\ref{kl}) we can calculate the total current at each interface using the expression:
\begin{equation}
eI_iR_N=\frac{\epsilon_i}{8\epsilon_{Th}}\int d\epsilon {\rm Tr\tau_3} \left[\check g_i,\check g\right]^K\; .
\label{currenti}
\end{equation}

In this case the GF inside S does not vary considerably and the Usadel equation (\ref{usadel}) can be integrated over space coordinates using the boundary conditions Eq. (\ref{kl}). In this way one  obtains a set of algebraic equations which can be written in a compact form
 \begin{equation}
 \left[ \check \Lambda, \check g\right]=0\label{central}
    \end{equation}
where 
\[
\check \Lambda=\sum_{i=L,R,S}\epsilon_i\check g_i+ \epsilon\tau_3+\check\Delta-\check \Sigma_{in}\; .
\]
Eq. (\ref{central}) is equivalent to the Nazarov's circuit theory equations \cite{nazarov_ct}, which were used in Refs. \cite{morten,morten08} for nonlocal transport calculations.
The solution for the R,A and K components of $\check g$ which satisfy Eq. (\ref{central}) and the normalization condition  can be formally be written as:
\begin{eqnarray}
\hat g^{R(A)}&=&\frac{\hat \Lambda^{R(A)}}{\sqrt{\hat \Lambda^{R(A)}\hat \Lambda^{R(A)}}}\label{sol1}\\
\hat g^K&=&\frac{\hat \Lambda^K-\hat g^R\hat \Lambda^K\hat g^A}{\sqrt{\hat \Lambda^R\hat \Lambda^R}+\sqrt{\hat \Lambda^A\hat \Lambda^A}}\label{sol2}
\end{eqnarray}
Substituting these expressions into Eqs. (\ref{delta}-\ref{current}) enable us to obtain numerically  the self-consistent  order parameter, the currents through the interfaces and the non-local voltage induced at the right electrode in the resistance measurement case.  
\begin{figure}[h]
\centerline{\includegraphics[scale=.35]{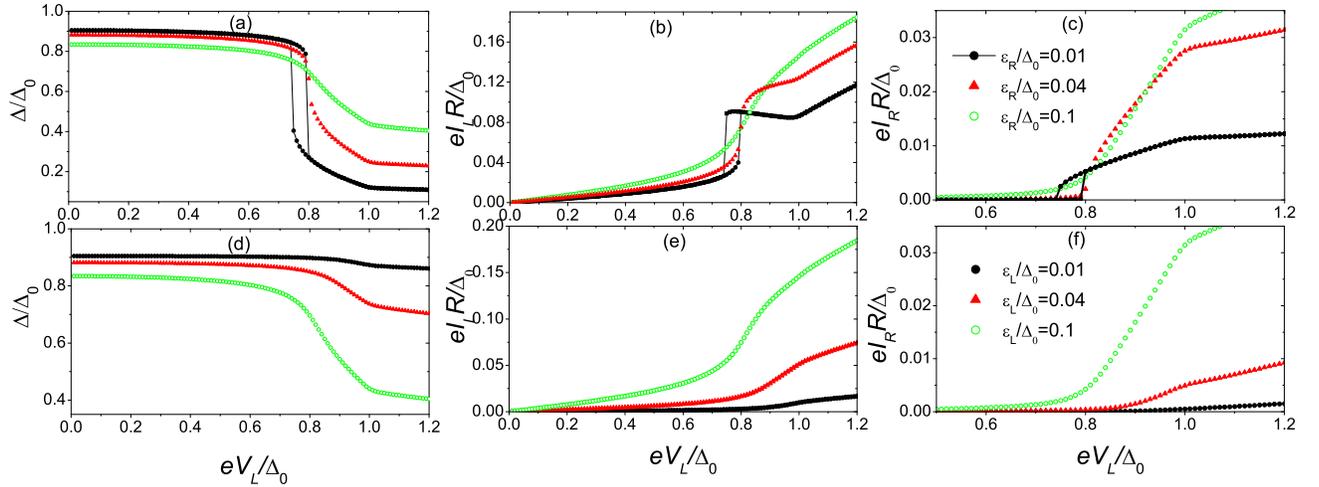}}
\caption{(Color online). The amplitude of the self-consistent order parameter, the current $I_L$ injected from the  
left normal electrode  and the current $I_R$ measured at the right electrode,  as a function of the voltage $V_L$. 
Panels (a)-(c) for  $\epsilon_S=0.2$, $\epsilon_L=0.1$, $T=0.01\Delta_0$ and different values of $\epsilon_R$. 
Panels (d)-(f) for $\epsilon_S=0.2$, $\epsilon_R=0.1$, $T=0.01\Delta_0$ and different values of $\epsilon_L$. 
We have defined $R=R_N\epsilon_{Th}/\Delta_0$. \label{fig-nl-cond-gap-current}}
\end{figure}

\section{Measurement of the non-local conductance}
In this section we consider the experimental set-up of Fig. 1a.  The left normal electrode is biased by a  voltage 
source at $V_L$. For a non-local conductance measurement we will assume that the $N_R$ is  grounded ($V_R=0$), and 
compute the current $I_R$ through the interface $S/N_R$ from Eqs. (\ref{currenti},\ref{sol1},\ref{sol2}) and the 
self-consistent order parameter $\Delta$ from Eq. (\ref{delta}). 
Before we address the non-local properties of the system let us discuss the results concerning local properties.
\begin{figure}[h]
\centerline{\includegraphics[scale=.35]{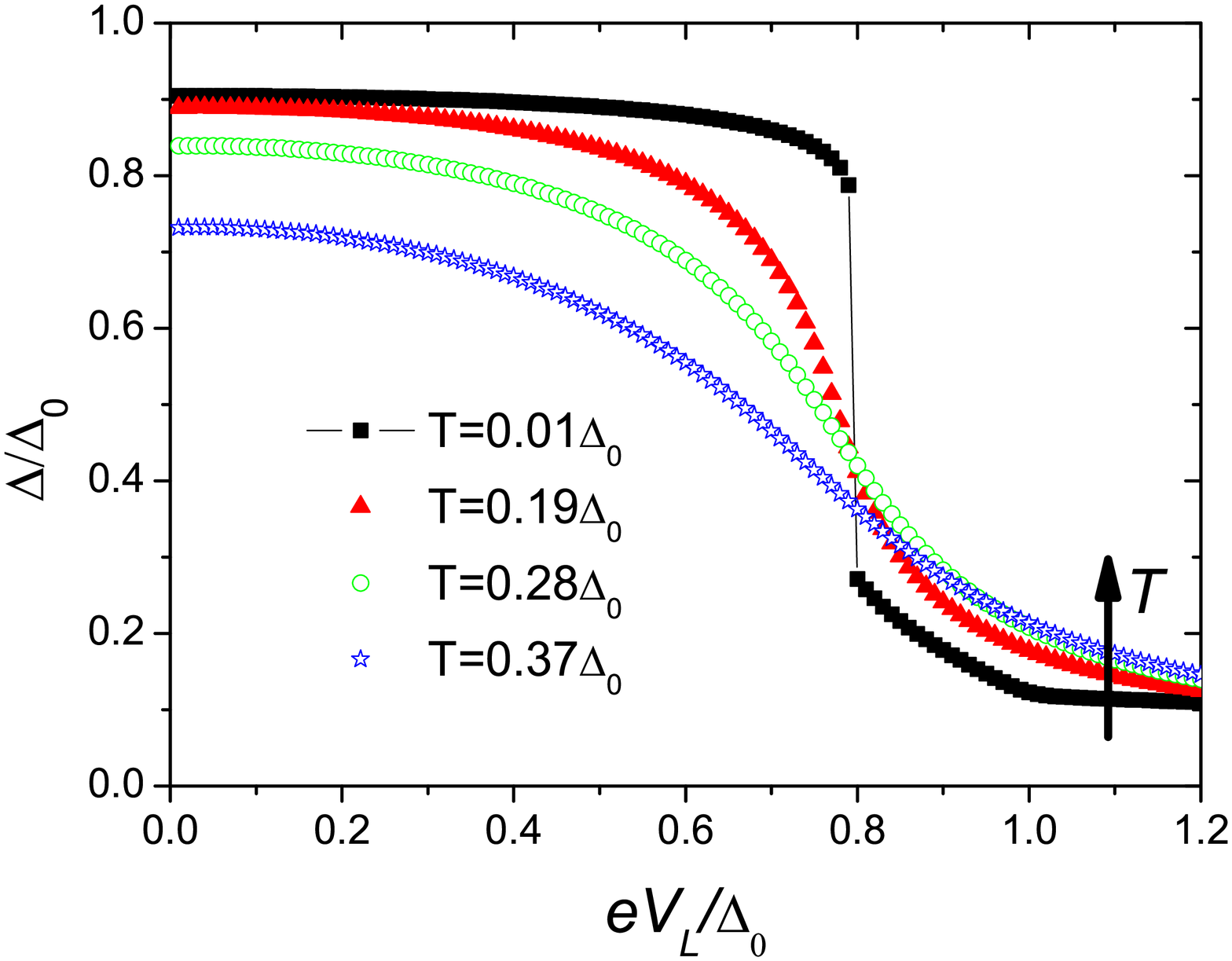}}
\caption{(Color online). The amplitude of the self-consistent order parameter as a function of the  voltage $V_L$, for different temperatures. Notice that in the range of temperatures shown and for large values of $V_L$ the gap is enhanced by increasing $T$.   \label{fig-nl-cond-gap-T}}
\end{figure}
In Fig. \ref{fig-nl-cond-gap-current}a we show the  amplitude of $\Delta$  as a function of the bias voltage $V_L$ for 
$\epsilon_S=0.2\Delta_0$, $\epsilon_L=0.1\Delta_0$ and three different  values of  $\epsilon_R=0.01-0.1$, corresponding  
to  transmission coefficients in the range $10^{-3}-10^{-2}$ (we assume that the length of the superconducting region is  
about $50 nm$). 
All energies are given in units of $\Delta_0$, which is the value of the order parameter in the bulk  at $T=0$.  
For a fixed low value of $V_L$, $\Delta$ is reduced by increasing the coupling with the right normal electrode, which is a 
consequence of  the inverse proximity effect. At some  value  $V_L^*\simeq 0.8\Delta_0$ of the order of the self-consistent $\Delta$, one can see an abrupt 
reduction of the self-consistent order parameter. For voltages  larger than $V_L^*$ the quasi-particle current through 
$S$ becomes considerably larger (Figs. \ref{fig-nl-cond-gap-current}b and \ref{fig-nl-cond-gap-current}c), {\it i.e.}  the system is driven out of equilibrium. 
An interesting  consequence of this non-equilibrium state for voltages $V_L>\sim V_L^*$ is the enhancement of the 
self-consistent gap by increasing the temperature, as shown in Fig. \ref{fig-nl-cond-gap-T}. 
This effect is related to the stimulation of superconductivity by quasiparticle currents in SIS systems, and was studied both theoretical\cite{zaitsev91} and experimentally\cite{blamire}.

From Fig. \ref{fig-nl-cond-gap-current}a one can also see that the suppression of $\Delta$ at $V_L^*$ becomes more 
abrupt the weaker the coupling with the right electrode is.  Figures \ref{fig-nl-cond-gap-current}b and 
\ref{fig-nl-cond-gap-current}c  also show the corresponding current at the left and right interfaces.  As expected the 
larger the resistance of the right interface (small $\epsilon_R$) the smaller the value of $I_R$.  
A strong non-equilibrium situation takes place when most of the current injected flows into the superconducting electrode, 
i.e. when $\epsilon_R$ is small enough (in our example $\epsilon_R=0.01$). In this case the gap becomes multivalued and 
this is reflected in the behavior of the currents $I_L$ and $I_R$. Multivalued solutions for the self-consistent gap were 
also found in Ref.\cite{zaitsev91} for SIS systems and recently in Ref.\cite{nazarov} for a NSN system.

Another interesting feature of this system is the existence of a region of voltages for which the  local conductance is negative (see Fig. \ref{fig-nl-cond-gap-current}b).  This  behavior was also obtained in Ref. \cite{volkov} for a system consisting of a superconducting link separating a normal and a superconducting electrode. Also in Ref.\cite{prb09} negative local conductance was obtained for a two dimensional ballistic superconductor attached to two normal electrodes. Notice however, that the part of the curve corresponding to a negative conductance would be not accessible in current biased experiments. 
If we now fix the value of $\epsilon_R$ at its maximum value $\epsilon_R=0.1$ and vary $\epsilon_L$,  we see  that even for the smallest coupling ($\epsilon_L=0.01\Delta_0$) the variation of the gap and  the currents is smooth and no signatures of multivalued solutions appears for this range of parameters (see  bottom row of Fig. \ref{fig-nl-cond-gap-current}).

\begin{figure}[h]
\centerline{\includegraphics[scale=.3]{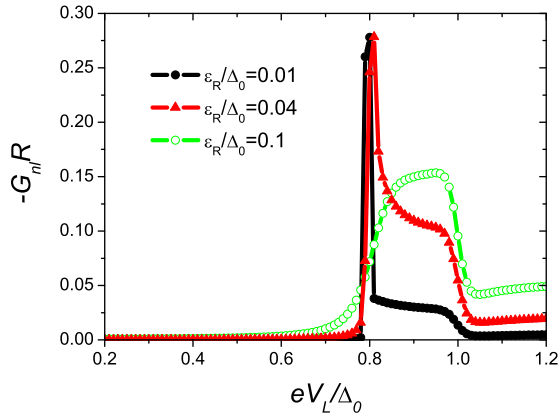}}
\caption{(Color online). The voltage dependence of the (negative) non-local conductance $G_{nl}$ normalized with respect to $R=R_N\epsilon_{Th}/\Delta_0$. We have chosen    $T=0.01\Delta_0$,  $\epsilon_S=0.2$, $\epsilon_L=0.1$, $T=0.01\Delta_0$ and different values of $\epsilon_R$ \label{fig-nl-cond}.}
\end{figure}

Let us now focus on the non-local transport and compute the  non-local conductance, which can  be obtained easily from the knowledge of $I_R$ (Fig. \ref{fig-nl-cond-gap-current}c and \ref{fig-nl-cond-gap-current}f) . It is given by the expression
\begin{equation}
G_{nl}=\frac{\partial I_R}{\partial V_L}\; .\label{grl}
\end{equation}
and shown in Fig. \ref{fig-nl-cond} as a function of the bias voltage $V_L$, for different values of the coupling energies $\epsilon_{L,R}$. In all cases  $G_{nl}$ is very small in the region of low voltages. This is in agreement with the zero non-local conductance obtained in the lowest order of  tunneling\cite{falci}, due to the cancellation of the EC and CAR processes. However, our results are in all order of tunneling and therefore a dominance of EC (negative $G_{nl}$) is obtained in accordance to Ref. \cite{regis}. 
 For voltages of the order of the self-consistent $\Delta$, the value of $G_{nl}$ becomes significant. However, no change of sign is observed. As mentioned in the introduction, non-interacting models do not exhibit a change of sign of the non-local conductance in three-terminal NSN structures. This can only be achieved when electron-electron interactions are taken into account\cite{natphys}.

\section{Measurement of the non-local resistance in a three-terminal device}
   
In a real experiment it is easier to measure a voltage rather than a current. Indeed  the  experiments of Refs. \cite{beckmann,chan,delft,beckmann2,basel} were performed  in (a) the current biased regime and  (b) instead of the current,  the non-local resistance (or voltage) was measured. 
Theoretically, it is not simple to impose a current bias. Therefore we will still work in the voltage biased case but  determine the induced non-local voltage and resistance imposing zero current at the $S/N_R$ interface. Thus,  all the  current injected from the left normal electrode flows into the superconducting electrode $S_e$ (see Fig. 1b).  The current at the right interface is given by ({\it cf.}  Eq.(\ref{currenti})):
\begin{equation}
eI_RR_{BR}=\frac{1}{8}\int d\epsilon {\rm Tr}\hat g^K(\epsilon)-\frac{1}{4}\int d\epsilon\nu_s(\epsilon)\left( \tanh(\frac{\epsilon+eV_R}{2T})-\tanh(\frac{\epsilon-eV_R}{2T})\right)\; ,
\label{currentr0}
\end{equation}
where $\nu_s(\epsilon)=(g^R-g^A)/2$ is the density of states of S and $R_{BR}$ is the $R$ barrier resistance per unit area. 
The first term in the r.h.s is proportional to the quantity $Q^*$ identified in the literature as  
the charge imbalance potential\cite{ci},  which appears due to a non-equilibrium distribution 
in the superconductor.  The second term is the usual quasiparticle current term. The voltage  $V_R$ is measured with 
respect to the ground (see Fig. 1b) and it is obtained by imposing $I_R=0$. 

\begin{figure}[h]
\centerline{\includegraphics[scale=.3]{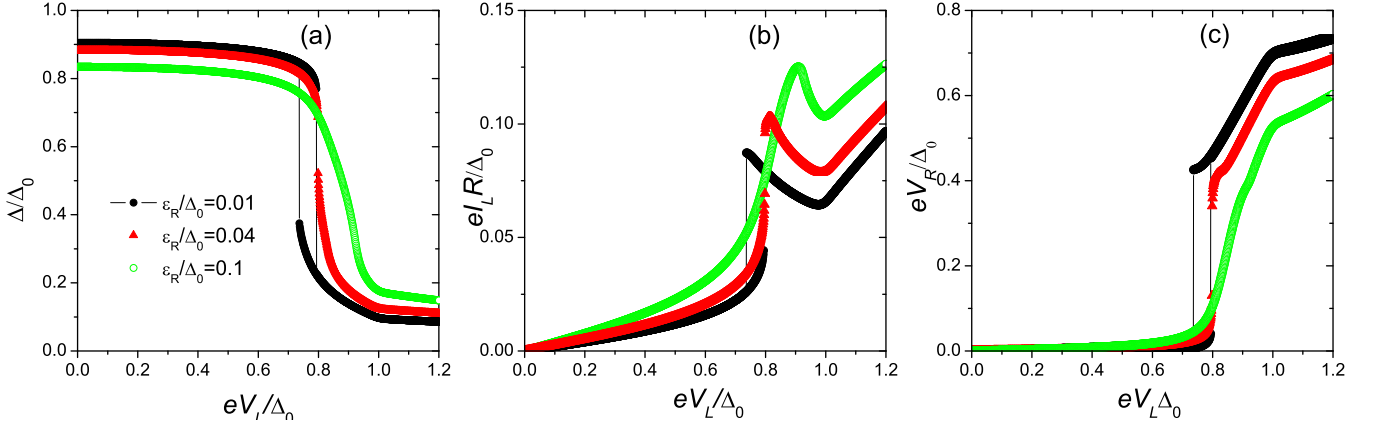}}
\caption{(Color online). The amplitude of the self-consistent order parameter (a), the current $I_L$ injected from the  left normal electrode (b)  and the induced voltage  $V_R$ measured at the right electrode (c),  as a function of the voltage $V_L$, for   $\epsilon_S=0.2$, $\epsilon_L=0.1$, $T=0.01\Delta_0$ and different values of $\epsilon_R$.\label{nl-res}}
\end{figure}

The results for $\Delta$, the injected current $I_L$ and the induced voltage $V_R$ as a function of the bias voltage $V_L$ are shown in Fig.  \ref{nl-res}, for fixed values of $\epsilon_{S,L}$ and different values of $\epsilon_R$ in the same range as in Fig. \ref{fig-nl-cond-gap-current}.
In the case of low values of $\epsilon_R$, the self-consistent gap has a very similar behavior as in the preceding section. However, for the largest value $\epsilon_R=0.1\Delta_0$ the suppression of $\Delta$ is larger as the one obtained  by imposing $V_R=0$. Notice also that the region of negative local conductance (Fig. \ref{nl-res}b), associated with the abrupt change of the order parameter 
appears now for all values of $\epsilon_R$.  It is clear that by having imposed $I_R=0$ all the current injected must flow through the $S/S_e$ interface and our  system behaves  similarly to the $N/S/S$ studied in Ref.\cite{volkov}.  
In Fig. \ref{nl-res}c we also show the  voltage $V_R$ induced in the right electrode calculated by equalizing 
(\ref{currentr0}) to zero. At low $V_L$ values the induced voltage $V_R$ is very small, but it experiences a jump at the value of $V_L$ where the self-consistent gap exhibits its maximal drop. 

\begin{figure}[h]
\centerline{\includegraphics[scale=.6]{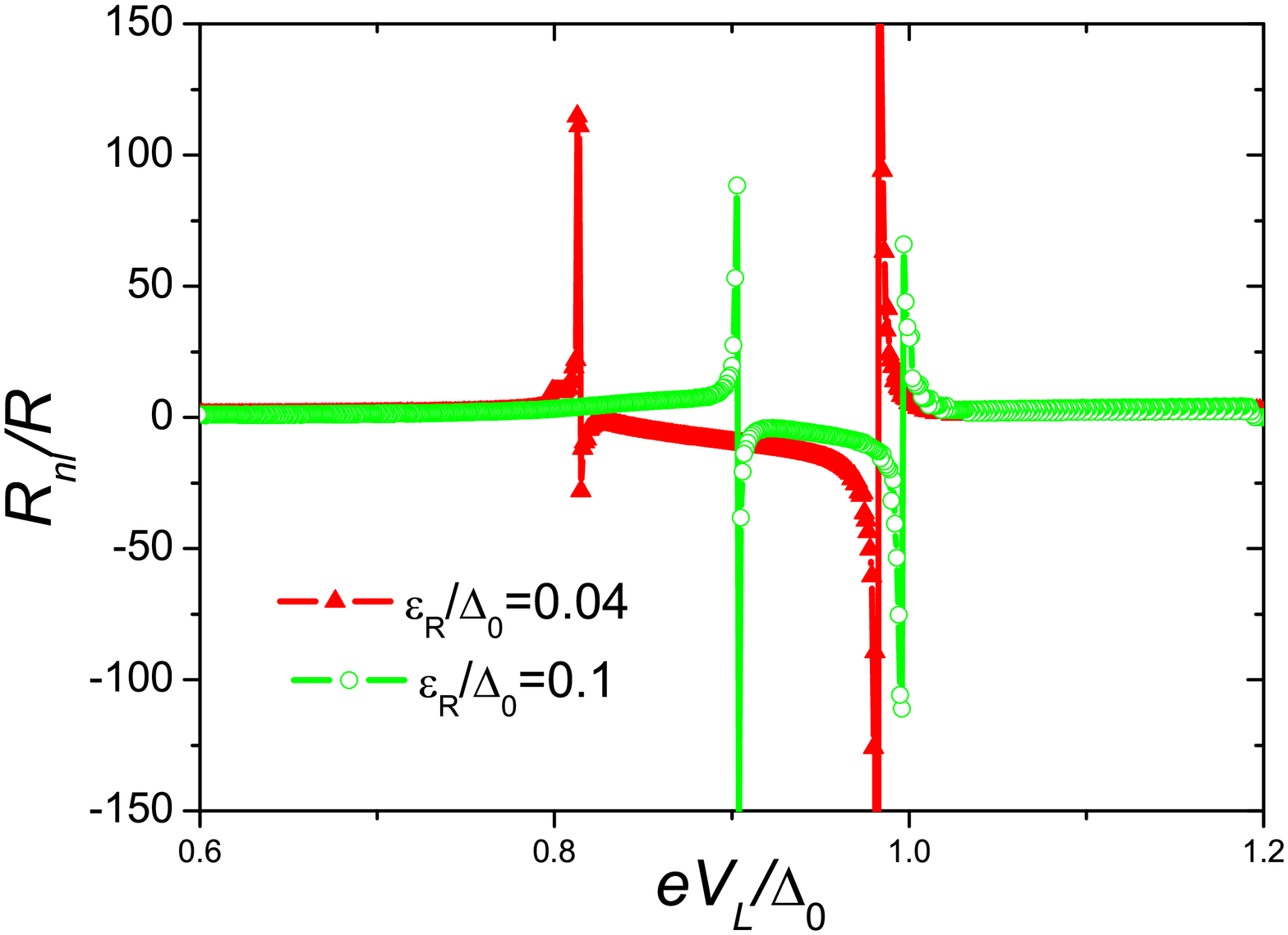}}
\caption{(Color online).  The voltage dependence of the non-local resistance calculated from Eq.(\ref{eqrab}) for $\epsilon_S=0.2\Delta_0$, $\epsilon_L=0.1\Delta_0$, $T=0.01\Delta_0$, and two values of $\epsilon_R$.\label{nl-res2}}
\end{figure}
We have now all quantities required to compute the non-local resistance which would be measured. This can be calculated 
from the expression:
\begin{equation}
R_{nl}=\frac{\partial V_{X}}{\partial I_L}=\frac{\partial V_{X}}{\partial V_L}\left(\frac{\partial I_L}{\partial V_L}\right)^{-1}\; .
\label{eqrab}
\end{equation}
The measured voltage $V_X$ depends on the experimental set-up. We are considering here the three-terminal structure of Fig. 1b, and  determining  $V_R$ respect to the ground. Thus in this case  $V_X=V_R$, {\it i.e.}  the one shown in \ref{nl-res}c. According to  Eq. (\ref{eqrab}) there are two factors
determining the  non-local resistance. One which is the inverse of the local conductance $G_{LL}^{-1}=dV_L/dI_L$ and which decrease by increasing $\epsilon_L$.  The second contribution is given by  $dV_R/dV_L$ which is nonzero only if a non-equilibrium distribution appears in the $S$ region and  is related to the charge imbalance term ({\it cf.} Eq. (\ref{currentr0})). 

In Fig. \ref{nl-res2} we show the dependence $R_{nl}(V_L)$  for two values of $\epsilon_R$ and $\epsilon_L=0.1\Delta_0$, 
$\epsilon_S=0.2\Delta_0$, $T=0.01\Delta_0$. The change of sign of $R_{nl}$ is a consequence of the negative local 
conductance which appears between the values  $V_L\simeq 0.8\Delta_0$ and $V_L\simeq\Delta_0$ (see Fig. \ref{nl-res}b). The position of the first peak of $R_{nl}$ is determined by the value of $V_L$ at which $\Delta$ drops substantially, while the position of second peak is determine by $\Delta_0$.   
As mentioned above,  in current bias experiments the curve $R_{nl}$ may look very different to the ones shown in Fig. \ref{nl-res2}, 
since the region of  negative  local conductance may not be observed.
 
In experiments as those of Refs\cite{beckmann,chan,basel} the current flows along a 
superconducting wire, while in our model it flows into the  reservoir $S_e$. For simplicity we have assume that the latter 
remains unaltered for all values $V_L$ considered here. In particular the value of the gap is the bulk BCS one for any 
value of $V_L$.  However, in the experiments,  when the current flowing through the wire reaches the critical value 
the superconducting gap is suppressed homogeneously in the region where the current is flowing.This  leads to the observation of only one peak in the non-local resistance$\cite{chan}$. In the next section we will model this situation. 
\begin{figure}[h]
\centerline{\includegraphics[scale=.3]{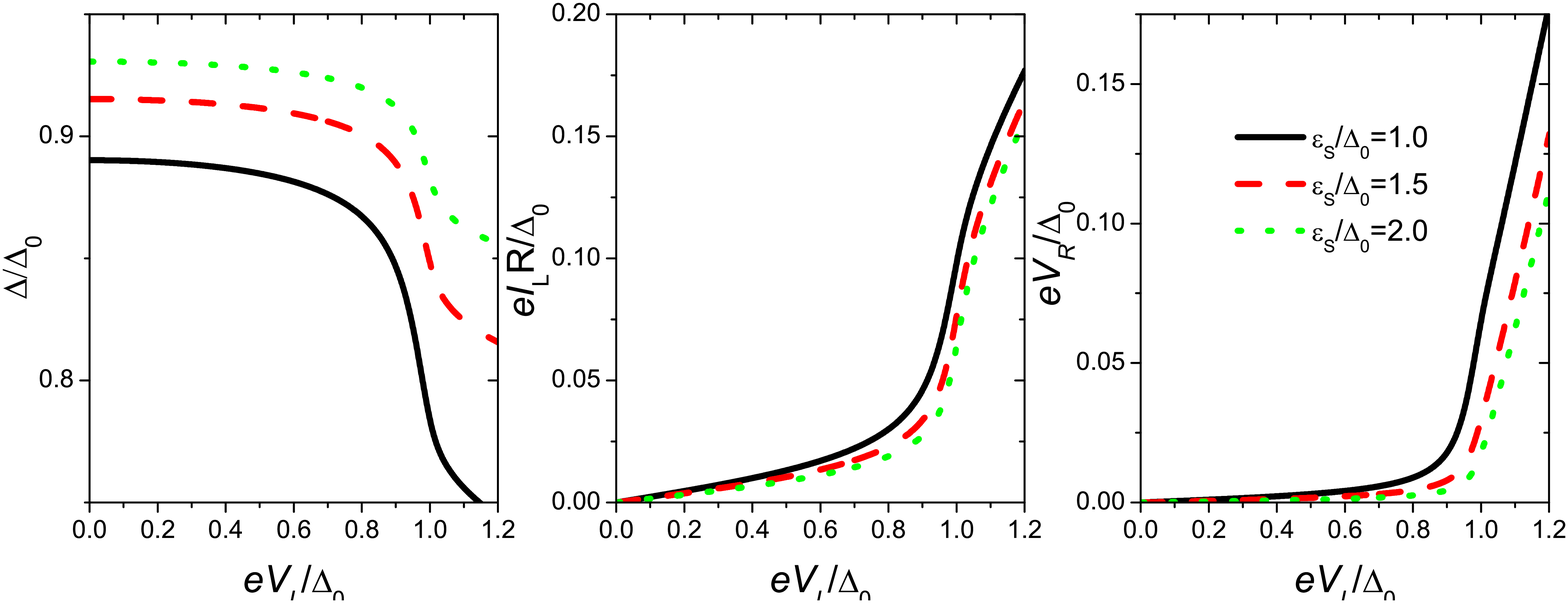}}
\caption{(Color online).  The amplitude of the self-consistent order parameter, the current $I_L$ injected from the  left normal electrode  and the induced voltage  $V_R$ measured at the right electrode,  as a function of the voltage $V_L$, for   $\epsilon_L=\epsilon_R=0.1$, $T=0.01\Delta_0$ and different values of $\epsilon_S$.\label{gap-large-es}}
\end{figure}
We should also emphasize that the change of sign of the non-local resistance obtained in Fig. \ref{nl-res2} is due to the 
fact that the local conductance $G_{LL}=dI_L/dV_L$ becomes negative for some values of $V_L$ 
({\it cf.} Fig. \ref{nl-res}b).   If  the coupling $\epsilon_S$ is large  enough,  the local conductance 
remains always positive and so the non-local resistance. This is shown  in Fig.\ref{gap-large-es}, where the amplitude of the 
self-consistency gap, the current injected and the voltage induced at the right electrode are plotted as a function of $V_L$, 
for $\epsilon_S=1,1.5$ and $2 \Delta_0$. If one compares these results with those obtained for a  smaller $\epsilon_S$ 
(Fig. \ref{nl-res}), one  sees that $\Delta$ is now only weakly suppressed and that the current $I_L$ increases monotonically. 
Thus the non-local resistance is always positive as it shown in Fig. \ref{rnl-large-es}. 
\begin{figure}[h]
\centerline{\includegraphics[scale=.35]{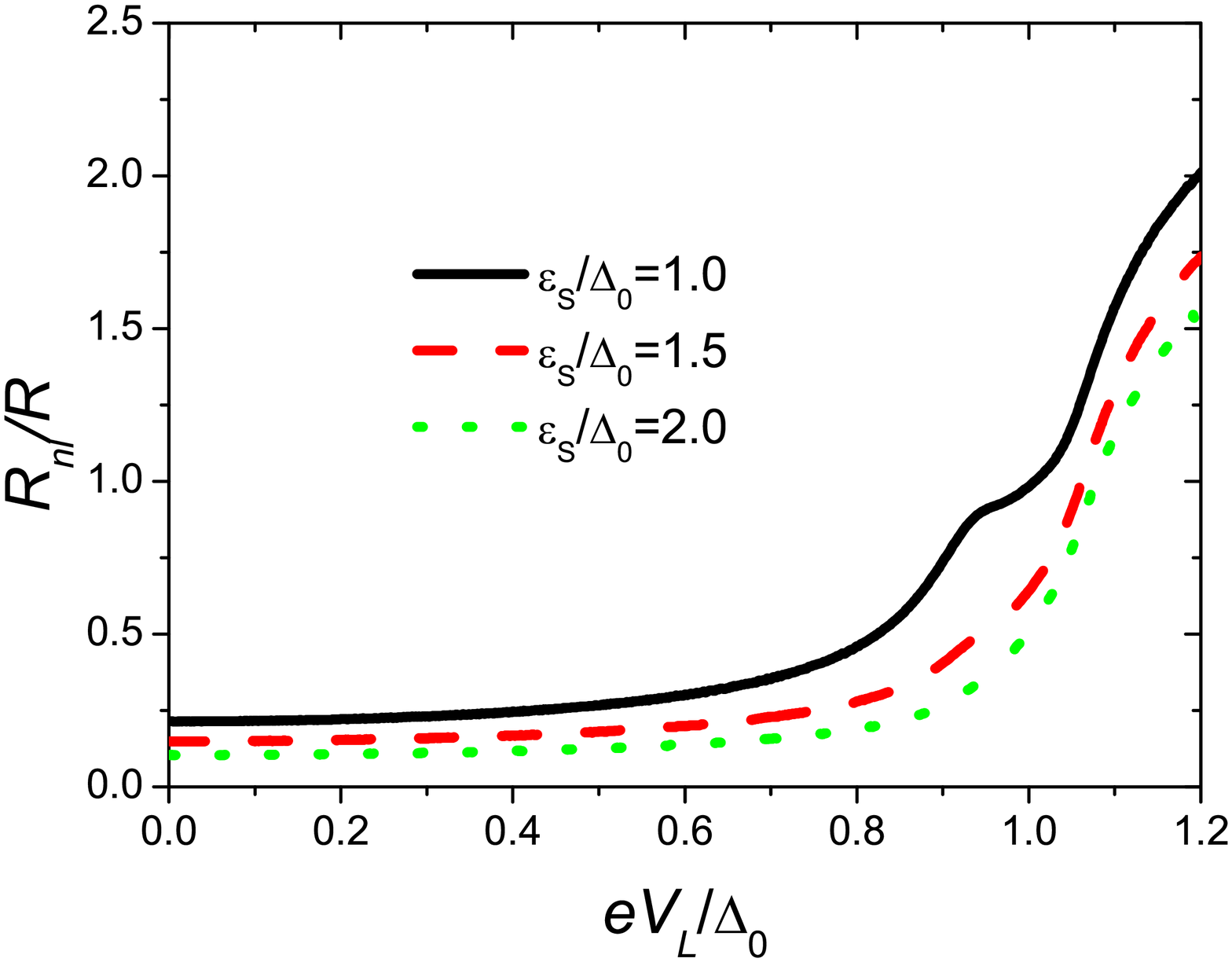}}
\caption{(Color online). The voltage dependence of non-local resistance for the same parameters as in Fig. \ref{gap-large-es} .\label{rnl-large-es}}
\end{figure}

\begin{figure}[h]
\centerline{\includegraphics[scale=.35]{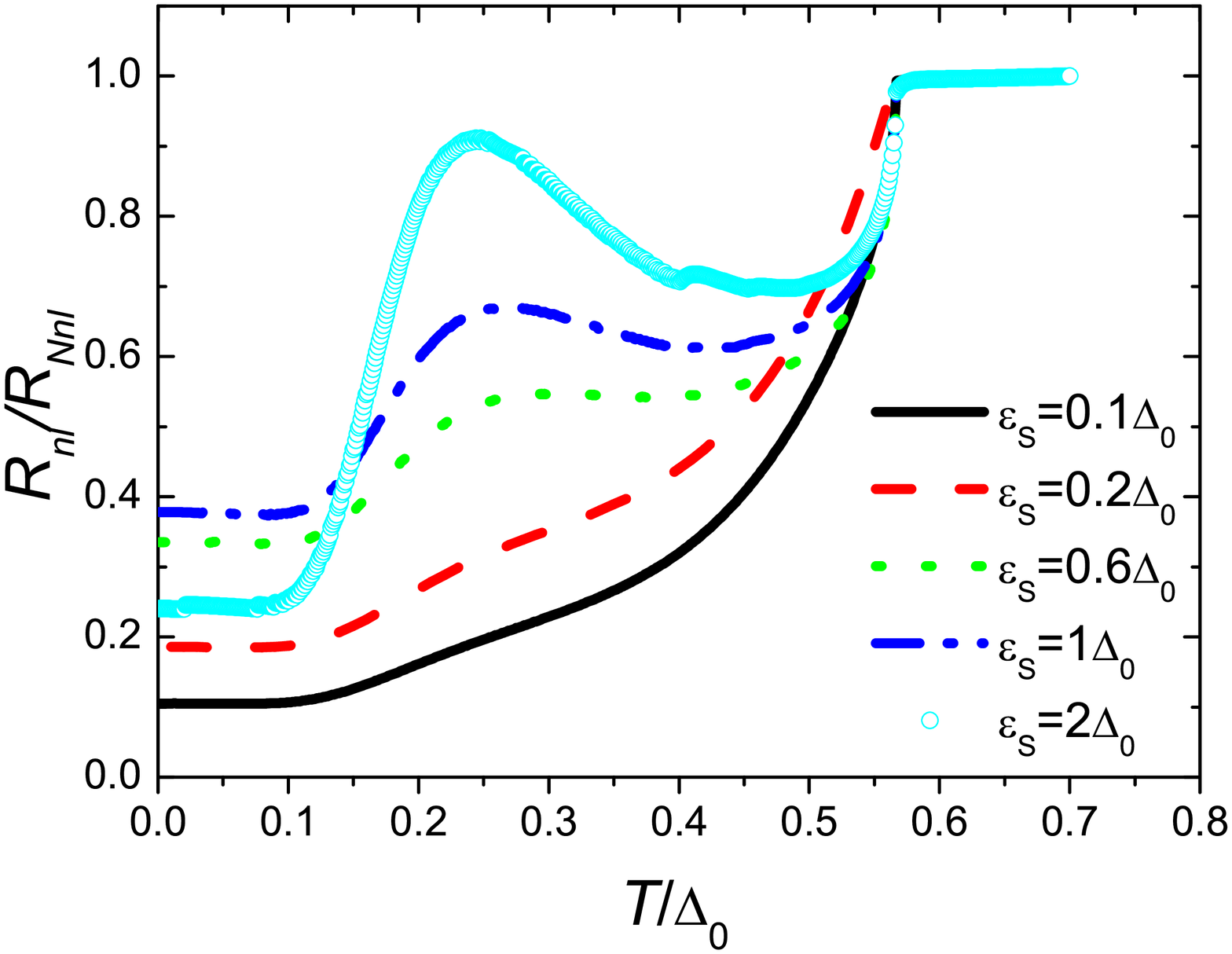}}
\caption{(Color online).  The temperature dependence of the zero bias non-local resistance measure in a three terminal device for different values of $\epsilon_S$. We have chosen  $\epsilon_L=\epsilon_R=0.1\Delta_0$ \label{nl-res-3T-tempdp}}
\end{figure}

Finally, we show in Fig. \ref{nl-res-3T-tempdp} the temperature dependence of the zero bias non-local resistance for 
different values of the coupling parameter $\epsilon_S$. We can see that while for the small values of $\epsilon_S$ 
the non-local resistance increases monotonously with the temperature, for larger values of $\epsilon_S$, $R_{nl}$ 
reaches a maximum value. In the latter case the charge imbalance effect becomes important and dominates over the local 
conductance factor for large temperatures. This behavior is in agreement with previous calculations of $R_{nl}$ in a 
superconducting quantum dot \cite{golubev}.  One could conclude as in Ref\cite{zaikin09} that the non-monotonic  
behavior of  $R_{nl}(T)$ is in qualitative agreement with the observations of Refs.\cite{beckmann,chan}. However, 
we hardly believe that. If this would be the case then one should obtain for the same parameter range a change of sign 
for $R_{nl}$ as a function of the injected current, as observed in  
the experiments\cite{chan}. On the contrary, Fig. \ref{rnl-large-es} clearly shows a monotonic increase of  $R_{nl}$ with the 
applied voltage.  In the next section we will show that the  change of sign of $R_{nl}$ observed in the experiment is due to the suppression of the superconductivity by the injection of a current. Also the peak of $R_{nl}$ observed as a function of temperature could be understood within this model.

\section{Four terminal  structure}

We now consider a situation closer to that of the experiments of Ref.\cite{chan,basel}, in which  the non-local resistance 
(voltage) has been measured in a multi-terminal  setup consisting  of a superconducting wire attached to several normal 
terminals. In these experiments the nonlocal voltage corresponds to the potential difference between 
one end of the superconducting wire and one of the normal leads. 
\begin{figure}[t]
\centerline{\includegraphics[scale=.35]{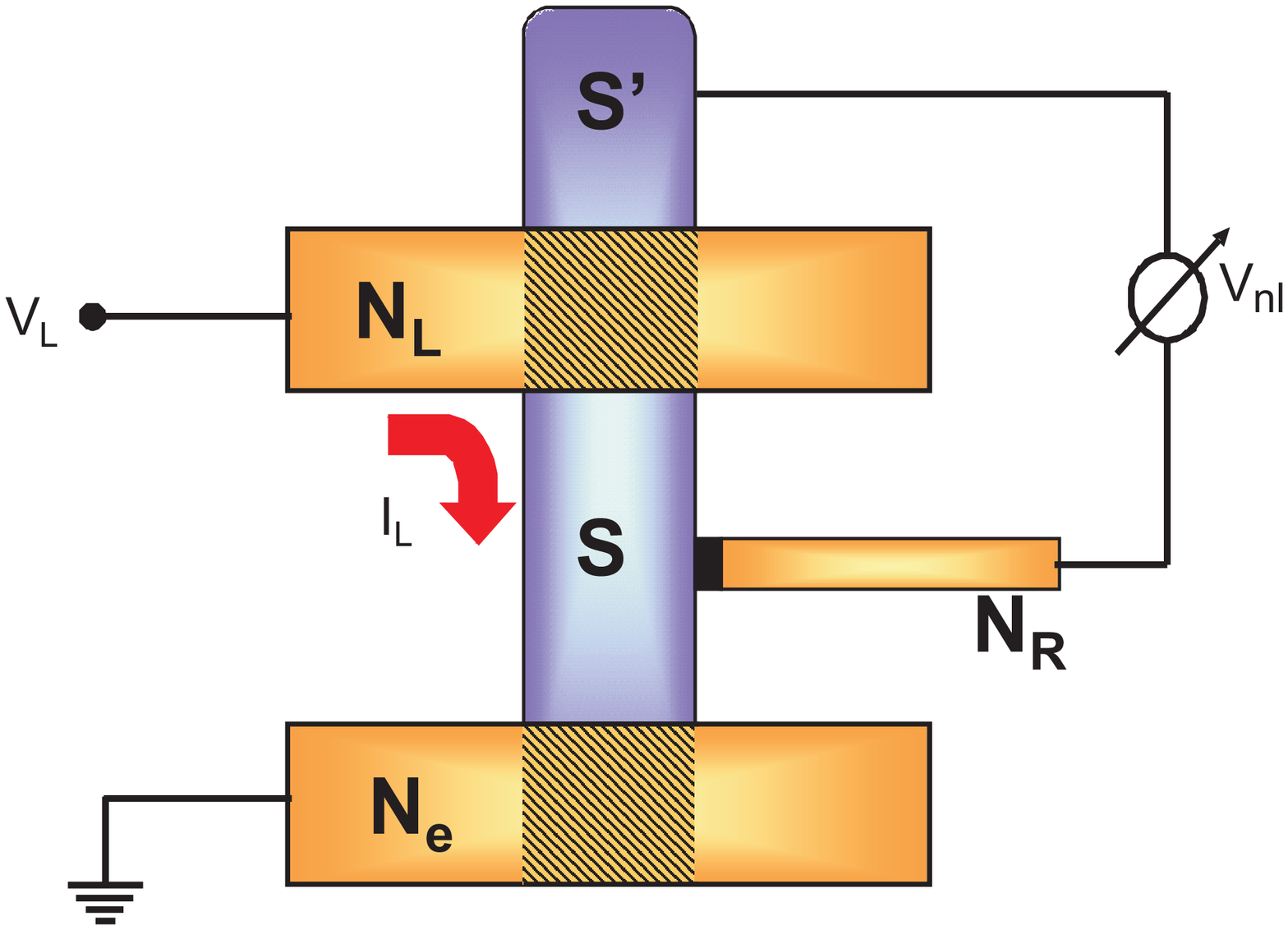}}
\caption{(Color online). Sketch of the 4-terminal structure under consideration.}
\label{geometry4T}
\end{figure}
We model these experimental situations as shown in Fig. \ref{geometry4T}. On the top of a superconducting wire we place 
two normal contacts. A current flows from the $N_L$ contact to the $N_e$ due to the bias voltage $V_L$ 
applied between the contacts. We are interested in the voltage difference measured between the end of the wire, which we 
denote by $S'$, and an additional normal contact $N_R$.  We use on purpose the same notation as in Fig. 1 in order 
 to use straightforwardly the expressions derived in section II. The only difference  is that  the drain electrode is now in 
the normal state and that we measure the voltage difference between the end $S'$ of the superconducting slab and the 
normal electrode $N_R$. The electrode $N_e$ is  grounded.  In order to compute the resistance measured between $S'$ 
and $N_R$ we proceed as in the last section, determining the Green functions from Eqs. (\ref{sol1}-\ref{sol2}) and  
the self-consistent gap. Now we impose that the currents through the $S/N_R$ and through the $S/S'$ interfaces are zero. 
The current trough the $S/S'$ interface can be written as the sum
\begin{equation}
I_{S/S'}=I_J+I_{qp}\;
\end{equation}
of the Josephson and the quasiparticle contribution respectively.  The first is given by the product of anomalous Green's 
functions in Eq. (\ref{current}) while the latter by the product of normal components. 
Since no current is flowing into  $S'$ we assume that there the Green's functions are those in equilibrium, with 
$|\Delta|$ equal to the self-consistent $|\Delta|$ in the S region at $V_L=0$, and a phase $\phi$ which is determined 
by imposing the condition of no current through the $S/S'$ interface. 
For voltages lower than a critical voltage $V_{L}^*\sim\Delta$ we always found a finite value of $\phi$. In this case the voltage 
$V'$ induced in $S'$ equals zero, and the measured $V_{nl}$ coincides with $V_R$ of the previous section. The voltage 
$V_{L}^*$  is the voltage at which the self-consistent gap vanishes, {\it i.e.} when the current flowing in $S$ reaches 
its critical value. For values of $V_L$ larger than this value the quasiparticle current becomes finite, and a voltage $V'$ 
is induced in $S'$. We compute it  from an expression obtained by equalizing Eq.(\ref{currenti}) to zero. Thus, the non-local resistance is 
given by Eq. (\ref{eqrab}) where now $V_X=V_R$ for $V_L<V_{L}^*$ and $V_X=V_R-V'$ for $V_L>V_{L}^*$. In Fig. \ref{nl-res4t} 
we show  the result of our calculation for $R_{nl}$ as a function of $V_L$. It shows a peak at $V_{L}^*$. Since the latter is of the order of $\Delta$, the peak is shifted to lower voltages by increasing the temperature (Fig. \ref{nl-res4t}) . This behavior 
is in agreement with the experimental observation of Ref. \cite{chan}, where the peak occurred at values of the bias 
current close to the value of critical current of the superconducting wire. The change of sign of $R_{nl}$ is related to the 
non-equilibrium situation created in $S$ by the injection of a current from $N_L$. 
We also show in the inset of Fig. \ref{nl-res4t} the temperature dependence of $R_{nl}$, which exhibits a pronounced peak for $V_L\simeq V_L^*$. This is  again  in agreement with the observations of Ref.\cite{chan}.  These results  demonstrate that the model studied here contains the main ingredients for describing experiments on non-local transport as the one of Ref.\cite{chan}. Within this model 
the change of sign of the non-local resistance has its origin in the deviation of the distribution function of the 
superconductor from the equilibrium one.  Notice, that as in Ref.\cite{chan}, the change of sign of $R_{nl}$ occurs 
at the critical current  which corresponds to the voltage $V_{L}^*$ in our model. 

 In other experiments\cite{delft,basel}, however,  the change of sign occurred at lower voltages. This discrepancy 
is at the moment not clarified and may be related to the inclusion of electron-electron interactions as proposed in 
Ref.\cite{natphys}.

\begin{figure}[h]
\centerline{\includegraphics[scale=.35]{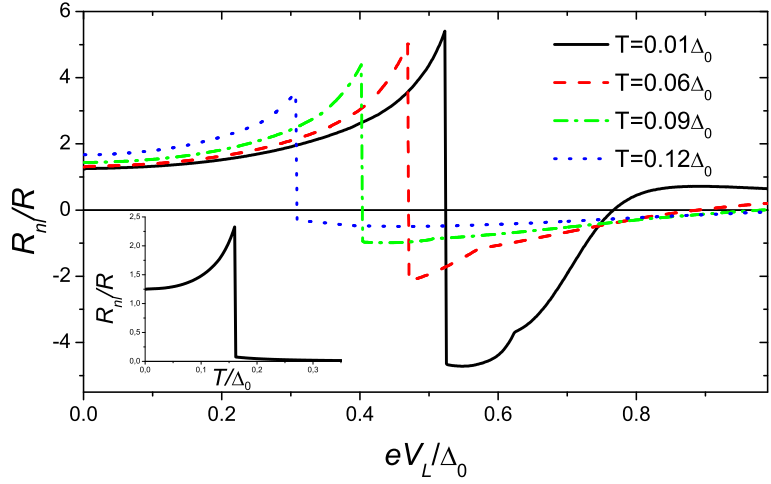}}
\caption{(Color online). The non-local resistance  measured in a four terminal structure as a function of the bias voltage $V_L$ for $\epsilon_L=\epsilon_R=0.1\Delta_0$,   $\epsilon_S=0.2$, and different values of the temperature. Inset: The non-local resistance as a function of the temperature at zero bias. \label{nl-res4t}}
\end{figure}


\section{Conclusions}

We have presented a self-consistent analysis of the transport properties of a structure consisting of a mesoscopic superconductor whose dimensions are smaller than the characteristic length $\xi_S$, attached to two normal and one superconducting terminals. We have analyzed two measurement methods:
one in which the detector (R) electrode is grounded and the leaking current is measured, and one in which the current through this lead is fixed
to zero and the induced voltage is measured. In both cases we observe that the self-consistent order parameter in the mesoscopic central region 
exhibits an abrupt drop at a certain voltage of the order of  the self-consistent $\Delta$. Associated to this drop the local differential conductance at the injector
lead (L) may become negative for certain values of the coupling parameters, resulting in a change of sign of the non-local resistance. As we stress
throughout this manuscript, this change of sign would not be related to a dominance of CAR over EC processes but to a non-equilibrium effect.
We have still described another mechanism for the appearance of negative non-local resistance which is probably most suitable for 
explaining the observations of Ref. \cite{chan}. This mechanism is applicable in a four terminal geometry and corresponds to the 
injection of large currents which may switch the superconducting region into the normal state. The observation of a change of sign in the
non-local signal at smaller bias and small transparencies like the ones reported in Refs. \cite{delft,basel} are certainly not possible to 
be explained with the theoretical model presented in this work and might be related to the influence of electron-electron interactions,
as already pointed out in Ref. \cite{natphys}. Further work for analyzing the combined effect of interactions and non-equilibrium
effects is under progress.

\section*{Acknowledgments}
We thanks R. M\'{e}lin for useful reading of the manuscript.
Financial support from Spanish MICINN under contracts FIS2005-06255 and FIS2008-04209 is acknowledged.
F.S.B. acknowledges funding by the Ram\'on y Cajal program.



\begin{thebibliography}{99}
\bibitem{beckmann} D. Beckmann, H.B. Weber and H. v. L\"ohneysen,
Phys. Rev. Lett. {\bf 93}, 197003 (2004).
\bibitem{delft} S. Russo, M. Kroug, T.M. Klapwijk and A.F. Morpurgo,
Phys. Rev. Lett. {\bf 95}, 027002 (2005).
\bibitem{chan} P. Cadden-Zimansky and V. Chandrasekhar,
Phys. Rev. Lett. {\bf 97}, 237003 (2006); 
P. Cadden-Zimansky, Z. Ziang and V. Chandrasekhar, New
J. Phys. {\bf 9}, 116 (2007).
\bibitem{beckmann2} D. Beckmann and H.  v. L\"ohneysen, Applied Physics A {\bf 89}, 603 (2007).
\bibitem{basel} A. Kleine, A. Baumgartner, J. Trbovic  and C. Sch\"oeneberger, cond-mat-01812.3553v2 (2009).
\bibitem{andreev} A.F. Andreev, Zh. Eksp. Teor. Fiz. {\bf 46},
1823 (1964) [Sov. Phys. JETP {\bf 19}, 1228 (1964)].
\bibitem{byers} J. M. Byers and M. E. Flatt\'e,  Phys. Rev. Lett.
  {\bf 74}, 306 (1995).
\bibitem{falci} G. Falci, D. Feinberg, and H. Hekking, Europhys. Lett.
{\bf 54}, 255 (2001).
\bibitem{regis} R. M\'elin and D. Feinberg, Phys. Rev. B {\bf 70}, 174509 (2004); S. Duhot and R. M\'elin, Eur. Phys. J. B {\bf 53}, 257 (2006); R. M\'elin, Phys. Rev. B {\bf 73}, 174512 (2006).
\bibitem{kalenkov} M.S. Kalenkov and A.D. Zaikin, JETP Lett. {\bf 87}, 140 (2008) [Pis'ma v ZhETF, {\bf 87}, 166 (2008)].
\bibitem{morten} J.P. Morten, A. Brataas and W. Belzig, Phys. Rev. B {\bf 74}, 214510 (2006).
\bibitem{golubov} A. Brinkman abd A.A. Golubov, Phys. Rev.B {\bf 74}, 214512 (2006).
\bibitem{golubev} D. S. Golubev and A. D. Zaikin, Phys. Rev. B {\bf 76}, 184510 (2007).
\bibitem{prb09} R. Melin, F. S. Bergeret, A. Levy Yeyati, Phys. Rev. B {\bf 79}, 104518 (2009).
\bibitem{zaikin09} D.S. Golubev, M.S. Kalenkov, A.D. Zaikin, arXiv:0904.3455. 
\bibitem{natphys} A. Levy Yeyati, F.S. Bergeret, A. Martin-Rodero and T.M. Klapwijk, Nature Phys. {\bf 3}, 455 (2007).
\bibitem{usadel} K.D. Usadel, Phys. Rev. Lett. {\bf 25}, 507 (1970). 
\bibitem{kl} M. Kupriyanov and V. F. Lukichev, Sov. Phys. JETP \textbf{67}, 1163 (1988).
\bibitem{feinberg} D. Feinberg, Eur. Phys. J. B \textbf{36}, 419 (2003). 
\bibitem{zaitsev91} A. V. Zaitsev, JETP Lett. {\bf 55},67 (1991).
\bibitem{blamire} M. G. Blamire, E. C. G. Kirk, J. E. Evetts and T. M. Klapwijk, Phys. Rev. Lett. {\bf 66}, 220 (1991).
\bibitem{nazarov_ct} Yu. V. Nazarov, Superlatt. Microstruct. {\bf 25}, 1221 (1999).
\bibitem{morten08} J. P. Morten, D. Huertas-Hernando, W. Belzig, and A. Brataas, Phys. Rev. B \textbf{78}, 224515 (2008).
\bibitem{nazarov} I. Snyman and Yu. V. Nazarov, Phys. Rev. B \textbf{79}, 014510 (2009).
\bibitem{zaikin07} M. S. Kalenkov and A. D. Zaikin, Phys. Rev. B \textbf{75}, 172503 (2007).
\bibitem{volkov} A. V. Zaitsev, A. F. Volkov, S. W. Bailey, and C. J. Lambert, Phys. Rev. B \textbf{60}, 3559 (1999).
\bibitem{ci} J. Clarke, Phys. Rev. Lett. \textbf{28}, 1363 (1972); M. Tinkham  and J. Clarke, Phys. Rev. Lett. \textbf{28}, 1366 (1972).
\end{thebibliography}
\end{document}